\documentclass[aps,prb,twocolumn,superscriptaddress,showpacs]{revtex4}
\usepackage{graphicx}
\usepackage{calc}
\usepackage{bm}

\begin{document}

\title{Manifestation of the bulk phase transition in the edge energy spectrum in a
two dimensional bilayer electron system}

\author{E.V.~Deviatov}
\email[Corresponding author. E-mail:]{dev@issp.ac.ru}
 \affiliation{Institute of Solid State
Physics RAS, Chernogolovka, Moscow District 142432, Russia}

\author{A.~W\"urtz}
\affiliation{Laboratorium f\"ur Festk\"orperphysik, Universit\"at
Duisburg-Essen, Lotharstr. 1, D-47048 Duisburg, Germany}

\author{A.~Lorke}
\affiliation{Laboratorium f\"ur Festk\"orperphysik, Universit\"at
Duisburg-Essen, Lotharstr. 1, D-47048 Duisburg, Germany}

\author{M.Yu.~Melnikov}
\affiliation{Institute of Solid State Physics RAS, Chernogolovka,
Moscow District 142432, Russia}

\author{V.T.~Dolgopolov}
\affiliation{Institute of Solid State Physics RAS, Chernogolovka,
Moscow District 142432, Russia}

\author{A.~Wixforth}
\affiliation{Institut f\"ur Physik, Universitat Augsburg,
Universitatsstrasse, 1 D-86135 Augsburg,  Germany}

\author{K.L.~Campman}
\affiliation{Materials Department and Center for Quantized Electronic
Structures, University of California, Santa Barbara, California
93106, USA}

\author{A.C.~Gossard}
\affiliation{Materials Department and Center for Quantized
Electronic Structures, University of California, Santa Barbara,
California 93106, USA}

\date{\today}

\begin{abstract}
We use a quasi-Corbino sample geometry with independent contacts
to different edge states in the quantum Hall effect regime to
investigate the edge energy spectrum of a bilayer electron system
at total filling factor $\nu=2$. By analyzing non-linear $I-V$
curves in normal and tilted magnetic fields we conclude that the
edge energy spectrum  is in a close connection with the bulk one.
At the bulk phase transition spin-singlet - canted
antiferromagnetic phase  $I-V$ curve becomes to be linear,
indicating the disappearance or strong narrowing of the  $\nu=1$
incompressible strip at the edge of the sample.
\end{abstract}


\maketitle

In a quantizing magnetic field, energy levels in a two dimensional
(2D) electron gas (2DEG) bend up near the edges of the sample,
forming edge states (ES) at the intersections with the Fermi
level. Electron transport through ES is responsible for many
transport phenomena in 2D, as it was firstly proposed by
B\"uttiker~\cite{buttiker} and further developed by Chklovskii
{\it et al.}~\cite{shklovsky} for interacting electrons. This ES
picture is  in good agreement with experimental
results~\cite{haug} on the transport both along ES and between
them.

Two principally  different sample geometries were applied for
transport investigations {\em between} different ES: (i) a
cross-gated Hall-bar~\cite{haug} and (ii) a split-gated
quasi-Corbino geometry~\cite{weiss,alida}. While measurements in a
Hall-bar geometry provide information on the equilibration length
between ES~\cite{haug,muller}, investigations in a quasi-Corbino
geometry are used to study the energy spectrum at the edge of a 2D
system~\cite{alida}. So far all experiments on the inter-edge
channel equilibration  have been performed on single-layer 2D
systems, despite bilayer electron systems also seem to be very
interesting.

 In the bulk of a bilayer system each Landau level is
split into four sublevels corresponding to the spin and
symmetric-antisymmetric splitting, which is caused by interlayer
tunnelling. In the simplest case of a weak Coulomb inter-layer
interaction the interplay between the symmetric-antisymmetric
splitting $\Delta_{SAS}$ and the Zeeman splitting is responsible
for the bulk properties of bilayer systems at the filling factor
$\nu=2$. $\Delta_{SAS}$ depends only on the electron
concentration, so at fixed total filling factor it is diminishing
with increasing magnetic field. In contrast, the Zeeman splitting
is proportional to the absolute value of the magnetic field. For
this reason, at total filling factor $\nu=2$ in low quantizing
magnetic fields two occupied energy levels are separated by a bare
Zeeman energy.
 These two occupied  levels belong to different spin
orientations so  that the system is in a spin-singlet state. The
excitation energy at filling factor $\nu=2$ is determined by the
next energy scale, i.e. the symmetric-antisymmetric splitting and
equals to $\Delta_{SAS}-g\mu B$. Increasing the magnetic field at
fixed total filling factor the excitation energy goes to zero. At
zero excitation energy $\Delta_{SAS}=g\mu B$ and a spectrum
reconstruction occurs: at higher fields $\Delta_{SAS}$ is the
minimal energy scale, so both the filled levels are at the same
spin orientation (in the field direction). The bilayer system is
called to be in a ferromagnetic state. This spin-singlet -
ferromagnetic phase transition can be driven also by an in-plane
field component at fixed normal magnetic field. Indeed, this is
only the Zeeman term which depends on the absolute value of the
field, while the others energy scales are determined by the normal
field component.

Regarding the  single-particle approximation (without significant
inter-layer interactions), while increasing the Zeeman splitting,
the bilayer system at the filling factor $\nu=2$ undergoes a
direct phase transition from a spin-singlet to a ferromagnetic
state~\cite{japan} at a critical magnetic field
$B_c=\Delta_{SAS}/g\mu$. However, in experiments with high
inter-layer Coulomb interaction~\cite{pellegrini,vadik} (the
distance between the layers is comparable to the magnetic length),
the transition point is significantly shifted into lower fields.
This was understood as a manifestation of many-body
effects~\cite{DasSarma,MacDonald,Iordanskii}. It was shown
theoretically that  the inter-layer Coulomb interaction shifts the
transition point  to a value $\mu
gB_c\approx\frac{\Delta_{SAS}^2}{E_c}$, where $E_c$ is the Coulomb
energy. At the field $B_c$ now a transition from the spin-singlet
to a novel canted antiferromagnetic state occurs. In this new
phase electron spins in both layers are canted from the field
direction due to the Coulomb interaction. This bulk phase was
experimentally investigated~\cite{vadik,pellegrini} and the
obtained results are in  good agreement with theoretical
predictions~\cite{DasSarma,MacDonald,Iordanskii}.

The situation at the sample edge is expected to be more
complicated. The ES  structure is determined by both the edge
potential and  the bulk  spectrum of a bilayer system. The latter
can be very complicated  even for the  simplest situation of total
filling factor $\nu=2$ in the
bulk~\cite{dqw,vadik,japan,pellegrini}. Moreover, the excitation
spectrum is strongly dependent on the local filling factor, which
varies widely at the sample edge. For these reasons even a
systematic of the excitations at the edge is unknown { \it ab
initio}.

Here we use a quasi-Corbino sample geometry to investigate the
edge spectrum of excitations at total filling factor $\nu=2$ in a
bilayer electron system in normal and tilted magnetic fields while
approaching the bulk phase transition from a spin-singlet to a
canted antiferromagnetic state. At the bulk transition point the
$I-V$ curve becomes to be linear indicating a strong narrowing of
the  incompressible strip between two ES.

Our bilayer structures are grown by molecular beam epitaxy on
semi-insulating GaAs substrate. The active layers form a 760~\AA\
wide parabolic quantum well. In the center of the well a 3
monolayer thick AlAs sheet is grown which serves as a tunnel
barrier between both parts on either side. The symmetrically doped
well is capped by 600~\AA\ Al$_x$Ga$_{1-x}$As ($x=0.3$) and
40~\AA\ GaAs layers. The symmetric-antisymmetric splitting in the
bilayer electron system as determined from far infrared
measurements and model calculations~\cite{hart} is equal to
$\Delta_{SAS}=1.3$~meV.

\begin{figure}
\includegraphics[width=\columnwidth]{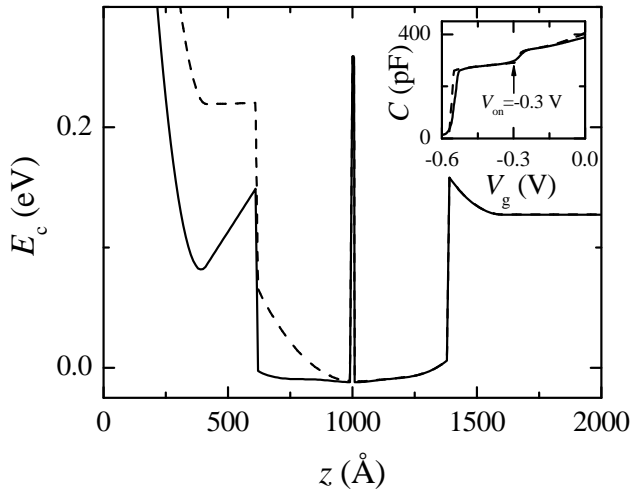}%
\caption{ Quantum well subband diagram at zero gate voltage (solid
line) and at twice smaller electron consentration (dashed line) as
calculated from the growth sequence of the structure (calculated
using Poisson-Schrodinger solver by G.~Snider). Inset shows the
capacitance of the sample in dependence on the gate voltage
calculated from the subband diagram (dashed) and measured in the
experiment (solid). The magnetic field is zero. \label{subband}}
\end{figure}

At zero gate voltage the quantum well is practically symmetric,
see Fig.~\ref{subband} (solid line). It contains
$4.2\times10^{11}$~cm$^{-2}$ electrons, which are distributed in
both parts of the well. Applying a negative voltage to the gate
makes the potential relief asymmetric (see Fig.~\ref{subband},
dashed line) indicating the depletion of the upper electron layer
at low enough voltages.

This is illustrated in the inset to Fig.~\ref{subband} where both
measured (solid) and calculated (dashed) capacitances are shown as
a function  of the gate voltage in zero magnetic field. At the
point of the abrupt changing of the capacitance (bilayer onset,
$V_{on}=-0.3$~V) electrons are leaving the top part of the well
and the distance between the gate and the 2D system is enlarged.

\begin{figure}
\includegraphics*[width=0.7\columnwidth]{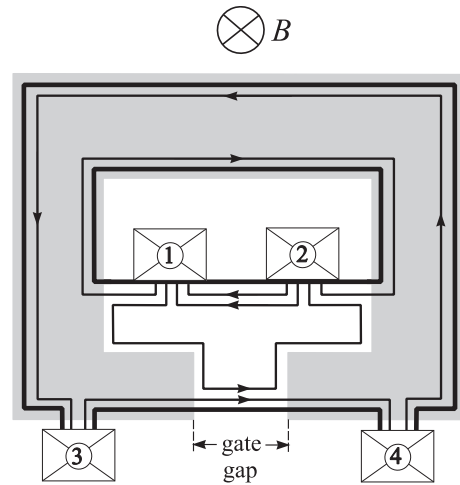}%
\caption{Schematic diagram of the pseudo-Corbino geometry.
Contacts are positioned along the etched edges of the ring-shaped
mesa (thick outline). The shaded area represents the
Schottky-gate. Arrows indicate the direction of electron drift in
the edge channels for the  filling factors $\nu=2$ in the ungated
regions and $g=1$ under the gate. \label{sample}}
\end{figure}

Samples are patterned in a quasi-Corbino geometry~\cite{alida}
(see Fig.~\ref{sample}). The square-shaped mesa has a rectangular
etched region inside. Ohmic contacts are made to the inner and
outer edges of the mesa (each of the contacts is connected to both
electron systems in the two parts of the well). The top gate does
not completely encircle the inner etched region but leaves
uncovered a narrow ($3\mu$m) strip (gate-gap) of 2DEG at the outer
edge of the sample.

At integer total filling factor $\nu=2$ edge channels are running
along the etched edges of the sample (see Fig.~\ref{sample}).
Depleting the electron system under the gate to a smaller integer
filling factor $g=1$ (as shown in the figure) one channel is
reflected at the gate edge and redirected to the outer edge of the
sample.  In the gate-gap region, ES originating from different
contacts run in parallel along the outer (etched) edge of the
sample, on a distance determined by the gate-gap width. Thus, the
applied geometry allows us to separately contact ES with different
spin and layer indexes and bring them into an interaction on a
controllable length.

In our experimental set-up one of the inner contacts is always grounded.
 We apply a dc current to one outer ohmic contact and measure the dc voltage
 drop between two remaining inner and outer contacts.
To increase the Zeeman splitting with respect to other energy
scales in our bilayer structure we apply an in-plane magnetic
field at fixed normal field by tilting the sample. Experiments are
performed at the temperature of 30 mK in  magnetic field up to
14~T.

\begin{figure}
\includegraphics[width=\columnwidth]{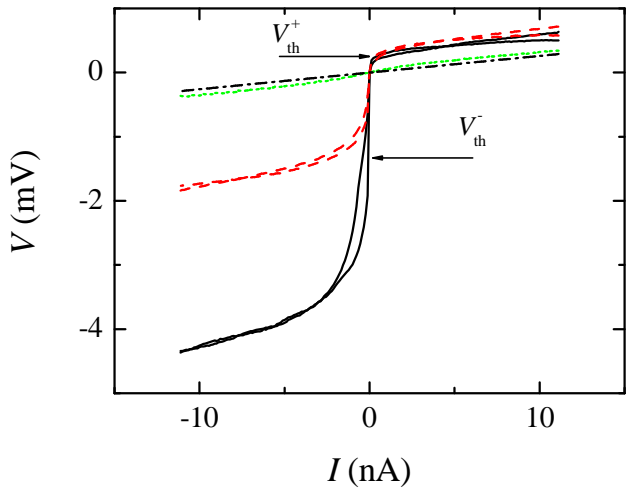}%
\caption{  $I-V$ curves for total filling factor $\nu=2$ and $g=1$
under the gate at different tilt angles. They are: $\theta=0$
(solid line), $\theta=30^\circ$ (dashed line), $\theta=45^\circ$
(dotted line). Dash-dot line depicts fully equilibrium $I-V$
curve, calculated from Landau-Buttiker formulas.  The normal
magnetic field is constant and equals to 8.7~T. \label{IV}}
\end{figure}

Measured $I-V$ curves are presented in Fig.~\ref{IV} for normal
and tilted magnetic fields for the filling factor  $\nu=2$ in the
gate-gap and $g=1$ under the gate.

In normal magnetic field the obtained $I-V$ curve is of a
diode-like form. It is non-linear and consists of two branches,
which starts from corresponding onset voltages - positive
$V_{th}^+$ and negative $V_{th}^-$ thresholds. In between these
thresholds a current is practically zero. The positive branch of
$I-V$ is close to be linear and characterizing by low resistance.
In contrast, the negative branch is strongly non-linear and of
higher resistance, see Fig.~\ref{IV}.

In normal magnetic field the positive threshold $V_{th}^+$ is
close to the bare Zeeman splitting (0.21~meV in the field of
8.7~T). The negative threshold  is one order of magnitude higher
($V_{th}^-$ is about 2~meV) and correspond to $\Delta_{SAS}$ in
our bilayer structure. In both cases it is a problem to estimate
the experimental accuracy - the exact value of the threshold
depends on the determination method. For example, the positive
threshold we can define either by an extrapolation from high
currents or as the voltage at which a significant current appears.
These values are slightly different, as can be seen from
Fig.~\ref{IV}. For the negative threshold the second method seems
to be more appropriate because of strong non-linear form of the
curve. Nevertheless all relevant energy scales in a bilayer system
(Zeeman splitting, symmetric-antisymmetric splitting and a
cyclotron splitting, which is about 15~meV here) are very
different, so it is easy to assign a threshold to the appropriate
spectral gap.

Both  threshold voltages are strongly dependent on the in-plane
field, see Fig.~\ref{IV}. They are diminishing with increasing
in-plane field and disappear at a tilt angle of $theta=45^\circ$.
A calculated $I-V$ trace for the case of  full equilibration
between two ES is also shown in Fig.~\ref{IV} (dash-dot) for the
comparison with $theta=45^\circ$ data. It can be seen that despite
disappearing of the threshold voltages at $theta=45^\circ$, the
experimental curve is still slightly non-linear.

The curves in Fig.~\ref{IV} are given for two sweep directions -
from positive to negative currents and vice versa. A small
hysteresis can be seen. It is a maximum in normal magnetic field,
becomes smaller at the tilt angle $\theta=30^\circ$ and disappears
at $\theta=45^\circ$. This hysteresis is a key feature for
transport between two spin-split ES~\cite{dixon,komiyama,DNP} -
for some electrons spin flip is accompanied by nuclear spin flop.
The hysteresis is an effect of the high nuclear relaxation time
(for a thorough discussion see Ref.~\onlinecite{DNP}).

\begin{figure}
\includegraphics[width=0.7\columnwidth]{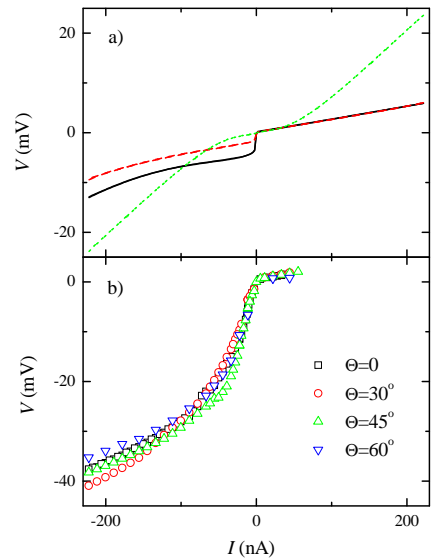}%
\caption{  $I-V$ curves for total filling factor $\nu=2$ and $g=1$
under the gate at different tilt angles in a wide current/voltage
range. a) The present bilayer system. The tilt angles are:
$\theta=0$ (solid line), $\theta=30^\circ$ (dashed line),
$\theta=45^\circ$ (dotted line). b) A single-layer
heterostructure, discussed in Ref.~\protect\onlinecite{DNP}. The
tilt angles are: $\theta=0$ (squares), $\theta=30^\circ$
(circles), $\theta=45^\circ$ (up triangles), $\theta=60^\circ$
(down triangles). \label{IVwide}}
\end{figure}

The dramatic influence  of the in-plane magnetic field on the
experimental $I-V$ traces can be also seen from Fig.~\ref{IVwide}
a). It demonstrates $I-V$ curves in a much wider current/voltage
range. The experimental non-linear $I-V$ curves are clearly
flattening when increasing the in-plane field. At a tilt angle of
$\theta=45^\circ$  even the curve shape is very different from the
normal field case.

The described behavior is totally different from that of a
single-layer structure, where no influence of the in-plane
magnetic field on the non-linear $I-V$ curves can be
observed~\cite{DNP}.  To demonstrate it in comparison with the
bilayer data we present in the Fig.~\ref{IVwide} b) $I-V$ curves
for a single - layer structure~\cite{DNP} at different tilt
angles. Because of high hysteresis in a single-layer case these
curves are obtained by waiting for 10 minutes at each point to
have time-independent $I-V$ curves. From both the values of the
thresholds and tilted field behavior we should conclude that
bilayer properties are important in the present experiment.

Using the gated part of the sample for magnetocapacitance
measurements we reproduced the previously obtained
results~\cite{dqw,vadik} on the bulk bilayer spectrum  at the
total filling factor $\nu=2$ in normal and tilted magnetic fields:
(i) In normal magnetic field the bulk activation energy , obtained
from the magnetocapasitance, is close to the single-particle
$\Delta_{SAS}$; (ii) While increasing the in-plane magnetic field
component the bulk bilayer system goes to the transition into the
canted antiferromagnetic phase.  This phase transition is
characterized by the appearance of a deep minimum in the
activation energy at the tilt angle of $\theta=45^\circ$.

\begin{figure}
\includegraphics*[height=0.4\textheight]{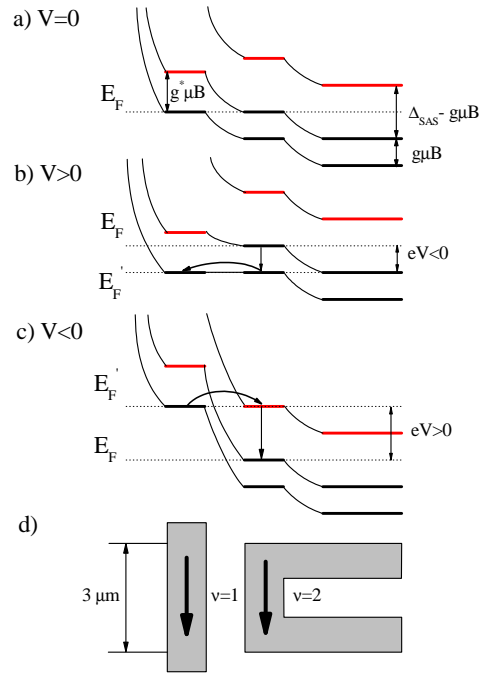}%
\caption{ Energy subband diagram of the sample edge in the
gate-gap for the filling factors $\nu=2$ and $g=1$. a) No voltage
$V$ applied between the inner and outer ES. b) $V>0$, in the
situation shown, the outer ES is shifted down in energy by $eV = -
\mid g \mid \mu_B B$. c) $V<0$, here the energy shift $eV>0$. d)
Sketch of the compressible strips in the gate-gap region. Arrows
in Figs.\protect\ref{ES},b) and \protect\ref{ES},c) indicate new
ways for electron relaxation opening up at high potential
imbalance. \label{ES}}
\end{figure}

Let us start the discussion from the case of normal magnetic
field. The bulk bilayer system at the filling factor $\nu=2$ is in
a spin-singlet state~\cite{dqw} and is  far from the phase
transition point. The energy structure in the bulk at the filling
factor $\nu=2$ can therefore be depicted in a single-particle
approximation  as two filled quantum levels under the Fermi
energy, separated by the Zeeman gap. The energy level structure is
depicted in Fig.~\ref{ES} a) in the gate-gap region. While
approaching the sample edge, two occupied energy levels bend up
because of rising the edge potential.

In our experimental geometry we independently contact inner
(always grounded) and outer ES. For this reason the measured
voltage drop $V$ is equal to the energy shift of the outer ES in
respect to the inner one, see Fig.~\ref{ES}, b)and c). For a
positive measured voltage $V>0$ the outer ES is shifted down in
energy by a value $eV<0$. It can be seen from Fig.~\ref{ES} b),
that at the value of the voltage $V=V_{th}^+= -g \mu B/e$ (where
$g$ is the bare g-factor, $e<0$ is the electron charge) the lowest
(the only occupied in both ES) energy level is flattened between
two ES, so that electrons can easily move from one ES into the
other. Thus, starting from the $eV_{th}^+=-g \mu B$ electrons can
be transferred between ES by vertical relaxation in the inner ES
between two spin-split levels and move along the lowest energy
level to the outer ES, see Fig.~\ref{ES}, b). This positive
voltage $V_{th}^+$ is characterized by a sharp rising of the
current, i.e. it is a threshold voltage for the positive branch of
the $I-V$ curve. It is a reason for the experimentally observed
$V_{th}^+$ to be close to the bare Zeeman gap. On the other hand,
a negative voltage shifts the outer ES up in energy. Electrons
always have to tunnel through the potential barrier from the outer
ES into the inner one. This tunnelling is only possible from the
occupied states  in the outer ES either into the empty states in
the inner ES or into the excited energy states in it. The later
process (with farther vertical relaxation in the inner ES to the
Fermi level) is more likely  for $eV>E_a$, where $E_a$ is the
energy of the first excited state. For this reason, experimental
$I-V$ traces change their slopes at $eV=E_a$, which we are
referring as the negative threshold voltage $V_{th}^-$. As we know
from bulk spectrum investigations, at the filling factor $\nu=2$
the first excited state is separated from the ground state by
$\Delta_{SAS}$. It is a reason for $V_{th}^-$ to be close to
$\Delta_{SAS}$ in normal magnetic field.

For an increase  of the in-plane magnetic field component a
Coulomb interaction becomes more and more important, so that the
simple single-particle picture described above is no longer
adequate. In the bulk the quantum levels are mixed into a new
ground state of the system, which is separated from the excited
state by a very low energy at the transition
point~\cite{DasSarma}. Approaching the sample edge, the electron
concentration is diminishing due to the edge potential. A local
filling factor is still $\nu=2$ before the inner ES and becomes to
be $\nu=1$ in between the inner and outer ES (see Fig.~\ref{ES}
d)). The energy structure at the edge is determined by a local
filling factor, so the system is still at $\nu=2$ ground state
before the inner ES and becomes to be at the $\nu=1$ ground state
between two ES.  In the inner ES the electron concentration is
changing from the value, corresponding to $\nu=1$ to the $\nu=2$
value.

 The electron system
in the vicinity of the $\nu=1$ incompressible strip can be
described as a $\nu=1$ ground quantum Hall state with some amount
of electron excitations (right side of $\nu=1$ strip in
Fig.~\ref{ES} d)), or as a $\nu=1$ ground quantum Hall state with
some amount of holes, on the opposite side of the $\nu=1$ strip.
It is the edge potential in the $\nu=1$ incompressible strip that
separates electrons and holes on both sides of the strip.
Consequently, the main result of our experiment should be
interpreted as a practical disappearance of this potential barrier
(or,possibly, of the incompressible $\nu=1$ strip itself) under
conditions of the canted antiferromagnetic phase in the bulk. In
these conditions electrons can freely move between ES without
spin-flips, so there is no reason both for the non-linear behavior
of experimental $I-V$ traces and for the hysteresis on them.

We used a quasi-Corbino sample geometry with independent contacts
to different edge states in the quantum Hall effect regime to
investigate the edge spectrum of a bilayer electron system at
total filling factor $\nu=2$. By analyzing non-linear $I-V$ curves
in normal and tilted magnetic fields we found that the edge energy
spectrum  is in a close connection with the bulk one. At the bulk
transition spin-singlet - canted antiferromagnetic phase the $I-V$
traces become to be linear indicating the disappearance  of the
potential barrier between $\nu=1$ ground state with some amount of
electron excitations and a $\nu=1$ ground state with some amount
of holes at the edge of the sample.

We wish to thank Dr. A.A.~Shashkin for help during the experiments
and discussions. We gratefully acknowledge financial support by
the Deutsche Forschungsgemeinschaft, SPP "Quantum Hall Systems",
under grant LO 705/1-2.  The part of the work performed in Russia
was supported by RFBR,  the programs "Nanostructures" and
"Mesoscopics" from the Russian Ministry of Sciences. V.T.D.
acknowledges support by A. von Humboldt foundation. E.V.D.
acknowledges support by Russian Science Support Foundation.

\end{document}